\documentstyle[12pt]{article}

\renewcommand{\theequation}{\thesection.\arabic{equation}}

\newlength{\extraspace}
\newlength{\extraspaces}
\newcounter{dummy}

\newcommand{\baa}{
\addtocounter{equation}{1}
\setcounter{dummy}{\value{equation}}
\setcounter{equation}{0}
\renewcommand{\theequation}{\thesection.\arabic{dummy}\alph{equation}}
\begin{eqnarray}
\addtolength{\abovedisplayskip}{\extraspaces}
\addtolength{\belowdisplayskip}{\extraspaces}
\addtolength{\abovedisplayshortskip}{\extraspace}
\addtolength{\belowdisplayshortskip}{\extraspace}}

\newcommand{\eaa}{
\end{eqnarray}
\setcounter{equation}{\value{dummy}}
\renewcommand{\theequation}{\thesection.\arabic{equation}}}

\newcommand{\be}{\begin{equation}
\addtolength{\abovedisplayskip}{\extraspaces}
\addtolength{\belowdisplayskip}{\extraspaces}
\addtolength{\abovedisplayshortskip}{\extraspace}
\addtolength{\belowdisplayshortskip}{\extraspace}}
\newcommand{\ee}{\end{equation}}

\newcommand{\ba}{\begin{eqnarray}
\addtolength{\abovedisplayskip}{\extraspaces}
\addtolength{\belowdisplayskip}{\extraspaces}
\addtolength{\abovedisplayshortskip}{\extraspace}
\addtolength{\belowdisplayshortskip}{\extraspace}}
\newcommand{\ea}{\end{eqnarray}}

\newcommand{\bd}{\begin{displaymath}
\addtolength{\abovedisplayskip}{\extraspaces}
\addtolength{\belowdisplayskip}{\extraspaces}
\addtolength{\abovedisplayshortskip}{\extraspace}
\addtolength{\belowdisplayshortskip}{\extraspace}}
\newcommand{\ed}{\end{displaymath}}

\newcommand{\ban}{\begin{eqnarray*}
\addtolength{\abovedisplayskip}{\extraspaces}
\addtolength{\belowdisplayskip}{\extraspaces}
\addtolength{\abovedisplayshortskip}{\extraspace}
\addtolength{\belowdisplayshortskip}{\extraspace}}
\newcommand{\ean}{\end{eqnarray*}}

\newcommand{\newsection}[1]{
\vspace{15mm}
\pagebreak[3]
\addtocounter{section}{1}
\setcounter{equation}{0}
\setcounter{subsection}{0}
\setcounter{footnote}{0}
\begin{center}
{\Large \thesection. #1}
\end{center}
\nopagebreak
\medskip
\nopagebreak}

\newcommand{\nonu}{\nonumber \\[.5mm]}

\newcommand{\deel}[2]{{\textstyle{#1 \over #2}}}
\newcommand{\hf}{{\textstyle{1\over 2}}}
\newcommand{\hs}{{\textstyle{1\over 6}}}
\newcommand{\hr}{{\textstyle{1\over 3}}}
\newcommand{\hh}[2]{{\textstyle{{#1}\over {#2}}}}

\newcommand{\ie}{{\it i.e.}}
\newcommand{\eg}{{\it e.g.\ }}
\newcommand{\re}{\mbox{I}\!\mbox{R}}
\newcommand{\lha}{\left[}
\newcommand{\rha}{\right]}

\newcommand{\uit}{\wedge}
\newcommand{\lie}[1]{{\bf #1}}
\newcommand{\dif}{\partial}
\newcommand{\dbar}{\bar{\dif}}
\newcommand{\abar}{\bar{A}}

\renewcommand{\ss}{\Sigma}
\newcommand{\tr}{\mbox{Tr}}
\newcommand{\del}{\delta}
\newcommand{\pri}{\Pi_i}
\newcommand{\prk}{\Pi_k}
\newcommand{\prid}{\Pi_i^{\dagger}}
\newcommand{\prkd}{\Pi_k^{\dagger}}
\newcommand{\vp}{\varphi}
\newcommand{\ro}{\rho}
\newcommand{\rb}{\bar{\rho}}
\newcommand{\mub}{\bar{\mu}}
\newcommand{\nub}{\bar{\nu}}

\newcommand{\var}[1]{{\del \over \del {#1}}}
\newcommand{\vars}[2]{{\del {#1} \over \del {#2}}}

\newcommand{\varsa}[1]{{\del {#1} \over \del \pri A}}

\newcommand{\varsabar}[1]{{\del {#1} \over \del \prk \abar}}

\newcommand{\mat}[9]{\left( \begin{array}{ccc}
				#1 & #2 & #3 \\
				#4 & #5 & #6 \\
				#7 & #8 & #9
                             \end{array} \right) }

\newcommand{\mats}[4]{\left( \begin{array}{cc}
				#1 & #2 \\
				#3 & #4
                             \end{array} \right) }

\newcommand{\cs}{Chern--Simons theory}

\newcommand{\np}[1]{Nucl. Phys. {\bf B#1}}
\newcommand{\cmp}[1]{Comm. Math. Phys. {\bf #1}}
\newcommand{\intmod}[1]{Int. Journal of Mod. Phys. {\bf A#1}}
\newcommand{\plb}[1]{Phys. Lett. {\bf B#1}}

\begin{document}
\addtolength{\baselineskip}{.7mm}

\thispagestyle{empty}
\begin{flushright}
{\sc THU}-91/18\\
10/91
\end{flushright}
\vspace{1.5cm}

\begin{center}
{\LARGE\sc{The Covariant $W_3$ Action}}\\[1.5cm]

\sc{Jan de Boer and Jacob Goeree}\\[8mm]
{\it Institute for Theoretical Physics\\[2mm]
University of Utrecht\\[2mm]
Princetonplein 5\\[2mm]
P.O. Box 80.006\\[2mm]
3508 TA Utrecht}\\[4cm]

{\sc Abstract}\\[1.5cm]
\end{center}

\noindent Starting with $Sl(3,\re)$ \cs\ we derive the
covariant action for $W_3$ gravity.

\vfill

\newpage

\newsection{Introduction}

Two dimensional gravity has been extensively studied during
the last few years. Three different approaches to the subject,
namely (i) study of the induced action of 2D gravity in both
the conformal (where it reduces to the Liouville action) as well
as the light cone gauge, (ii) the discretized
approach of the matrix models, and (iii)
topological gravity, have all been very powerful
(at least for $c < 1$), giving equivalent results.

Higher spin extensions of 2D gravity can also be studied using
the above methods. These theories are commonly denoted as
theories of $W$ gravity. Especially $W_3$ gravity in the light
cone gauge has been the subject of many recent research
\cite{kj1,kj2,pope,hull}. In this paper we will also concern ourselves
with the study of $W_3$ gravity, but from a different angle.
Believing that the `$W_3$ moduli space' is somehow related to the
moduli space of flat $Sl(3,\re)$ bundles, we will study $W_3$
gravity starting from $Sl(3,\re)$ \cs\, whose classical
phase space is the space of flat $Sl(3,\re)$ bundles.

Our analysis resembles the one in \cite{herman}.
In this reference H. Verlinde showed how the physical state
condition in $Sl(2,\re)$ \cs\ can be reduced to the conformal
Ward identity, giving as a by-product the fully covariant action
of 2D gravity. We will start with $Sl(3,\re)$ \cs, and derive
the covariant action for $W_3$ gravity. It will turn out that
this action describes $Sl(3,\re)$ Toda theory coupled to a
`$W_3$ background,' confirming general beliefs.
Although we restrict ourselves to the case of $W_3$ in this
paper, we believe that many of our results can be generalized.
This will be reported elsewhere \cite{jj}.
\newsection{\cs}

\cs\ on a three manifold $M$ is described by the action
\be \label{eq:cs}
S = \frac{k}{4\pi i} \int_{M} \tr(A\uit dA+\deel{2}{3}A\uit A\uit A),
\ee
where the connection $A$ is a one form with values in the Lie
algebra \lie{g} of some Lie group $G$, and $d$ denotes the
exterior derivative on $M$. In this paper $M$ will be
of the form $M=\ss \times \re$, $\ss$ being a
Riemann surface, for which
$A$ and $d$ can be decomposed into space and time components,
\ie\ $A=A_0dt+\tilde{A}$, with
$\tilde{A}=\tilde{A}_zdz+\tilde{A}_{\bar{z}}d\bar{z}$,
and $d=dt\dif / \dif t+\tilde{d}$.
Rewriting the action as
\be \label{eq:cs2}
S = \frac{k}{4\pi i} \int dt \int_{\ss}
\tr(\tilde{A}\uit \dif_t\tilde{A} + 2A_0(\tilde{d}\tilde{A}
+\tilde{A}\uit \tilde{A})),
\ee
we recognize that $A_0$ acts as a Lagrange multiplier which
implements the constraint $\tilde{F}=\tilde{d}\tilde{A} +
\tilde{A}\uit \tilde{A}=0$. Furthermore, we deduce from this
action the following non-vanishing Poisson brackets
\be \label{eq:poisbracket}
\{ \tilde{A}^a_{\bar{z}}(z),\tilde{A}^b_{z}(w) \}=
\frac{2 \pi i}{k}\eta^{ab}\del(z-w),
\ee
where $\tilde{A}_{z}=\sum_a \tilde{A}^a_{z}T^a$, with
$\tr(T^aT^b)=\eta^{ab}$.

Upon quantizing the theory we have to replace the above
Poisson bracket by a commutator, and we have to choose a
`polarization.' This simply means that we have to divide
the set of variables
$(\tilde{A}^a_{z},\tilde{A}^a_{\bar{z}})$ into two subsets.
One subset will contain fields $X_i$ and
the other subset will consist of derivatives
$\var{X_i}$, in accordance with (\ref{eq:poisbracket}).
The choice of these subsets is called
a choice of polarization. Of course we also have to incorporate
the Gauss law constraints $\tilde{F}(\tilde{A})=0$. Following
\cite{herman,moore,witten1} we will impose these constraints
{\em after} quantization. So we will first consider a `big'
Hilbert space obtained by quantization of
(\ref{eq:poisbracket}), and then select the physical states $\Psi$
by requiring $\tilde{F}(\tilde{A})\Psi=0$.

In \cite{herman} it was shown that these
physical state conditions for $Sl(2,\re)$ \cs\ with a
certain choice of polarization are equivalent to the conformal
Ward identities satisfied by conformal blocks in Conformal Field
Theory (CFT). More precisely, it was shown that two of the three
constraints in $\tilde{F}(\tilde{A})\Psi=0$ could be explicitly
solved, leaving one constraint which is equivalent to
the conformal Ward identity. In this paper we will generalize
these results to the case of $Sl(3,\re)$ with a choice of
polarization that leads to the Ward identities of the $W_3$
algebra \cite{sammy}. (A different choice of polarization
leading to the related $W_3^2$ algebra was made in \cite{bilal2}.)
To explain our strategy, we will in the next section first
reconsider the case of $Sl(2,\re)$ \cs.

\newsection{$Sl(2,\re)$}

In order to understand how one obtains the Virasoro Ward
identity from $Sl(2,\re)$ \cs , let us first recall how one can
in general obtain Ward identities from zero-curvature
constraints. Given operator valued connections $A$ and
$\bar{A}$\footnote{Note that from now on $\tilde{A}$
will be denoted as $A$.},
the zero curvature condition reads:
\be
F\,\Psi =\, :\dif \abar - \dbar A + [A, \abar]:\Psi=0. \label{zerof}
\ee
Here the dots denote normal ordering, which
simply amounts to putting all $\delta/\delta X$ to the right.
(Here $X$ is some arbitrary field.)
If we take\footnote{In the following $\var{X}$ should in fact
read $\deel{2\pi}{k}\var{X}$, for all fields $X$.}:
\be
A=\mats{0}{1}{\frac{\delta}{\delta \mu}}{0}, \label{asimple}
\ee
and put $\abar_{12}=\mu$, we can solve for the remaining
components of $\abar$, if we require that the curvature operator
must have the form
\be
F=\mats{0}{0}{\ast}{0}. \label{Ftype}
\ee
Equation (\ref{zerof}) reduces to one equation which is
precisely the Virasoro Ward identity for $c=6k$
\be
\lha (\dbar-\mu\dif-2(\dif\mu))\var{\mu}+\hf\dif^3\mu\rha\Psi=0.
\ee
If we start with $Sl(2,\re)$ \cs\ and pick a polarization, then
$A$ and $\abar$ consist of three fields and their variational
derivatives. On the other hand, the Virasoro Ward identity contains
just one field. So in order to obtain the Virasoro Ward identity
as zero curvature constraint of \cs , we must introduce two
extra degrees of freedom without changing the contents of the zero
curvature constraints. We can in this case simply introduce
extra degrees of freedom by performing a gauge transformation.
Under such a gauge transformation the curvature transforms
homogeneously:$\,\, F \rightarrow g^{-1}Fg$, and $F=0$ will
still give the Virasoro Ward identity. If we
require $F$ to remain of type (\ref{Ftype}),
$g$ must be of the form
\be
g=\mats{g_{11}}{0}{g_{21}}{g^{-1}_{11}}. \label{gtype}
\ee
We can parametrize $g$ via a Gauss
decomposition:
\be g=\mats{1}{0}{\chi}{1} \mats{e^{\phi}}{0}{0}{e^{-\phi}}, \ee
and we find that $F\rightarrow Fe^{2\phi}$. The gauge
transformed $A,\abar$ are
\ba
A^g & = & \mats{\chi+\dif\phi}{e^{-2\phi}}{e^{2\phi}(
\dif\chi-\chi^2+\var{\mu})}{-\chi-\dif\phi}, \label{atra} \\
\abar^g & =  & \mats{\hf\dif\mu+\chi\mu+
\dbar\phi}{\mu e^{-2\phi}}{e^{2\phi}(
\mu\var{\mu}-\hf\dif^2\mu-\mu\chi^2-\chi\dif\mu+
\dbar\chi)}{-\hf\dif\mu-\chi\mu-\dbar\phi}.
\label{abartra} \ea
If we pick the following polarization:
\ba
A & = &
\mats{A_{11}}{A_{12}}{\var{\abar_{12}}}{-A_{11}}, \label{polara} \\
\abar & = &
\mats{-\hf\var{A_{11}}}{\abar_{12}}{-\var{A_{12}}}{\hf\var{A_{11}}},
\label {polarabar}
\ea
and let $F(A,\abar)$ act on a wavefunction $e^S \Psi[\mu]$, where $S$
still has to
be determined, we find that
$F(A,\abar)e^S\Psi=0 \Leftrightarrow e^SF(A',\abar
')\Psi=0$, with\footnote{Here we ignored terms of the type
$\frac{\delta^2 S}{\delta X\delta X'}$,
that involve delta-function type
singularities that have to be regularized somehow. Therefore
the validity of our discussion will be limited to the semi-classical
level. In the full quantum theory
we expect corrections to the expressions given in this paper.}
\ba
A' & = & \mats{A_{11}}{A_{12}}{\vars{S}{\abar_{12}}
	 +\var{\abar_{12}}}{-A_{11}}, \label{aprime} \\[.5mm]
\abar' & = & \left( \begin{array}{cc}
	     -\hf\vars{S}{A_{11}}-\hf\var{A_{11}} & \abar_{12}\\[.7mm]
-\vars{S}{A_{12}} -\var{A_{12}} & \hf\vars{S}{A_{11}}+\hf\var{A_{11}}
\end{array} \right).
\label{abarprime}
\ea

To proceed, suppose that we want to match
(\ref{aprime}) and (\ref{abarprime}) with
(\ref{atra}) and (\ref{abartra}). This gives
\ba
A_{11} & = & \chi+\dif\phi, \nonumber \\
A_{12} & = & e^{-2\phi}, \label{parama} \\
\abar_{12} & = & \mu e^{-2\phi}. \nonumber
\ea
Using these expressions, we can express
the variational derivatives with respect to
$A$ and $\abar$ in terms of those with respect to
$\chi$, $\phi$ and $\mu$. One
finds:
\ba
\var{A_{11}} & = & \var{\chi}, \nonumber \\
\var{A_{12}} & = & -\hf e^{2\phi}
\left(\dif \var{\chi} + \var{\phi} -2\mu\var{\mu}
\right), \label{paramadif} \\
\var{\abar_{12}} & = & e^{2\phi} \var{\mu}. \nonumber
\ea
Quite remarkably, the terms containing $\var{\mu}$ in (\ref{aprime}) and
(\ref{abarprime}) agree precisely with
those in (\ref{atra}) and (\ref{abartra}).
As $\Psi$ depends only on $\mu$,
we can omit the $\var{\phi}$ and $\var{\chi}$
terms in (\ref{abarprime}).
In conclusion, we see that (\ref{aprime}) and
(\ref{abarprime}) are exactly
identical to (\ref{atra}) and (\ref{abartra}) if the
following relations hold:
\ba
\vars{S}{A_{11}} & = & -\dif\mu-2\chi\mu-2\dbar\phi, \nonumber \\
\vars{S}{A_{12}} & = & e^{2\phi} \left(\hf\dif^2\mu
+\mu\chi^2+\chi\dif\mu-\dbar\chi \right), \label{sequ} \\
\vars{S}{\abar_{12}} & = & e^{2\phi} (\dif\chi-\chi^2). \nonumber
\ea
Before continuing, we will first rewrite
these expressions in a form that is more
suitable for generalizations to other cases.
Let $G$ be the subgroup of $Sl(2,\re)$
consisting of all $g$ of type (\ref{gtype}), and let
$\Lambda=\mats{0}{1}{0}{0}$, which is equal to
(\ref{asimple}) up to terms containing $\var{\mu}$.
Furthermore, let $(\abar_{0})_{12}=\mu$ and let $\abar_{0}$ be such that
$F(\Lambda,\abar_{0})$ is of type (\ref{Ftype}).
For this $\abar_{0}$ one finds
$F(\Lambda,\abar_{0})=\mats{0}{0}{-\hf\dif^3\mu}{0}$.
To rewrite the polarizations,
define projections $\prk$ and $\pri$ on the Lie algebra $sl(2,\re)$ via
\ban
\prk \mats{b}{a}{c}{-b} & = & \mats{0}{a}{0}{0}, \\
\pri \mats{b}{a}{c}{-b} & = & \mats{b}{a}{0}{-b}.
\ean
The polarization (\ref{polara}) and (\ref{polarabar})
is such that the fields are
in $\pri A$ and $\prk \abar$, and that the derivatives are in
$\prid\abar$ and $\prkd A$, where $\prid=1-\prk$ and $\prkd=1-\pri$. Now
(\ref{atra}) and (\ref{abartra}) read,
up to terms containing $\var{\mu}$,
$A=g^{-1}\Lambda g + g^{-1}\dif g$
and $\abar=g^{-1}\abar_0 g + g^{-1}\dbar g$.
Therefore, (\ref{parama}) can be compactly formulated as
\ba
\pri A & = & \pri(g^{-1}\Lambda g +g^{-1}\dif g), \label{afields} \\
\prk \abar & = & \prk(g^{-1}\abar_0 g +g^{-1}\dbar g),
\label{abarfields}
\ea
and equations (\ref{sequ}) become
\ba
-\varsa{S} & = & \prid(g^{-1}\abar_0 g +g^{-1}\dbar g),
\label{Sequ1} \\
\varsabar{S} & = & \prkd(g^{-1}\Lambda g + g^{-1}\dif g).  \label{Sequ2}
\ea
Surprisingly, these equations can be
integrated\footnote{The proof of this fact will be given elsewhere
\cite{jj}.} to give
\be
S=\deel{k}{2\pi}\int d^2z\,\,\tr(\prkd A \prk \abar)
-\deel{k}{2\pi}\int d^2z\,\,
\tr(\Lambda \dbar g g^{-1})-\Gamma_{WZW}[g],
\label{chiralaction}
\ee
where $\Gamma_{WZW}$ is the Wess-Zumino-Witten action
\be
\Gamma_{WZW}[g]=\deel{k}{4\pi}\int d^2z\,\,
\tr(g^{-1}\dif g g^{-1}\dbar g)
-\deel{k}{12\pi}\int_B\,\,\tr(g^{-1}dg)^3 \label{WZW}.
\ee

For $Sl(2,\re)$ we find that
\be
S=\deel{k}{2\pi}\int d^2z\,\,\lha
\mu\dif\chi-\mu\chi^2-2\dbar\phi\chi-\dif\phi\dbar\phi \rha,
\ee
which indeed solves (\ref{sequ}).
To make contact with the results of \cite{herman}
we have to redefine $\phi \rightarrow -\hf \vp$, and
$\chi \rightarrow \hf \omega +\hf \dif \vp$. Then the action becomes
$S=S_{0}(\omega,\vp,\mu)+S_{L}(\vp,\mu),$
where
\be
S_{0}=-\deel{k}{4\pi} \int d^2z\,\,\lha \hf \mu\omega^2
-\omega(\bar{\dif}\vp-\dif \mu -\mu \dif \vp ) \rha, \label{so}
\ee
and $S_L$ is a chiral version of the Liouville action:
\be
S_L=\deel{k}{4\pi}\int d^2z\,\,\lha
\hf \dif \vp \bar{\dif} \vp + \mu (\dif^2 \vp
-\hf (\dif \vp)^2) \rha. \label{sl}
\ee
The actions (\ref{so}), (\ref{sl}) are precisely
the same as the ones found in \cite{herman}.
Altogether we have now shown how the
Virasoro Ward identity follows from $Sl(2,\re)$ \cs .
In general, the procedure consists of three steps:
(i) pick a polarization and parametrization of the
components of $A$ and $\abar$,
(ii) move $A$ and $\abar$ through a term of the form
$e^S$ and (iii), perform a gauge transformation.
In the next section we will apply
these steps to the case of $Sl(3,\re)$ \cs .

\newsection{$Sl(3,\re)$}

The Ward identities of the $W_3$-algebra can obtained in
the same way as the Virasoro Ward
identity was obtained in the previous section. We start with
\be
A=\mat{0}{1}{0}{0}{0}{1}{\var{\nu}}{\var{\mu}}{0}, \label{defaa}
\ee
and put $\abar_{13}=\nu$ and $\abar_{23}=\mu$.
The remaining components of $\abar$ are fixed
by requiring $F$ to be of the form
\be
F=\mat{0}{0}{0}{0}{0}{0}{F_{31}}{F_{32}}{0}.   \label{fform}
\ee
As was shown in \eg \cite{kj1}, $F_{31}$ and $F_{32}$
are directly related to the $W_3$-Ward
identities. The subgroup $G$ of $Sl(3,\re)$ that
preserves this form of $F$ consists of all $g\in
Sl(3,\re)$ satisfying $g_{13}=g_{23}=0$.
To parametrize these $g$, we will again use a Gauss
decomposition:
\be
g=\mat{1}{0}{0}{\phi_1}{1}{0}{\phi_3}{\phi_2}{1}
\mat{e^{\alpha}}{0}{0}{0}{e^{\beta-\alpha}}{0}{0}{0}{e^{-\beta}}
\mat{1}{\chi}{0}{0}{1}{0}{0}{0}{1}. \label{defgg}
\ee
Under a gauge transformation $F\rightarrow g^{-1}Fg$ we find
\ba
F_{31}' & = & e^{\alpha+\beta}F_{31}+\phi_1e^{\alpha+\beta}F_{32}, \\
F_{32}' & = & \chi e^{\alpha+\beta}F_{31}+
(\chi\phi_1e^{\alpha+\beta}+e^{2\beta-\alpha})F_{32},
\ea
which clearly shows that $F\Psi[\mu,\nu]=0
\Leftrightarrow (g^{-1}Fg)\Psi[\mu,\nu]=0$. The
polarization we choose is such that it is
invariant under the subgroup $G$ of $Sl(3,\re)$:
\ba
A & = & \mat{A_++A_-}{A_{12}}{A_{13}}{A_{21}}{-2A_+}{A_{23}}{
\var{\abar_{13}}}{\var{\abar_{23}}}{A_+-A_-}, \label{defpolara} \\
\abar & = & \mat{-\hs\var{A_+}-\hf\var{A_-}}{-\var{A_{21}}}{\abar_{13}}{
-\var{A_{12}}}{\hr\var{A_+}}{\abar_{23}}{-\var{A_{13}}}{-\var{A_{23}}}{
-\hs\var{A_+}+\hf\var{A_-}}, \label{defpolarabar}
\ea
and the projections $\pri$ and $\prk$ are given by
\ba
\prk \mat{a+b}{c}{d}{e}{-2a}{f}{g}{h}{a-b} & = &
\mat{0}{0}{d}{0}{0}{f}{0}{0}{0}, \\
\pri \mat{a+b}{c}{d}{e}{-2a}{f}{g}{h}{a-b} & = &
\mat{a+b}{c}{d}{e}{-2a}{f}{0}{0}{a-b}.
\ea
The matrices $\Lambda$ and $\abar_0$ are also
found completely analogously to the
$Sl(2,\re)$-case, one simply requires $F(\Lambda,\abar_0)$
to be of type (\ref{fform}), to find:
\ba
\Lambda & = & \mat{0}{1}{0}{0}{0}{1}{0}{0}{0}, \\
\abar_0 & = & \mat{\dif\mu+\hh{2}{3}\dif^2\nu}{\mu+\dif\nu}{\nu}{
-\dif^2\mu-\hh{2}{3}\dif^3\nu}{-\hr\dif^2\nu}{\mu}{
\dif^3\mu+\hh{2}{3}\dif^4\nu}{-\dif^2\mu-
\hr\dif^3\nu}{-\dif\mu-\hr\dif^2\nu}.
\ea
It is straightforward to read of
the explicit expressions for $\pri A$ and
$\prk \abar$ from (\ref{afields}) and (\ref{abarfields}). They are:
\ba
A_+ & = & \hf \left( \phi_1 -\phi_2+(\phi_1^2-\phi_3)\chi
e^{2\alpha-\beta}-\chi\dif\phi_1
e^{2\alpha-\beta} + \dif (\alpha-\beta) \right) \nonumber \\
A_- & = & \hf \left( \phi_1 +\phi_2+(\phi_1^2-\phi_3)\chi
e^{2\alpha-\beta} -\chi\dif\phi_1
e^{2\alpha-\beta}+ \dif (\alpha +\beta) \right) \nonumber \\
A_{12} & = & (2\phi_1-\phi_2)\chi + e^{\beta-2\alpha} +
\chi^2(\phi_1^2-\phi_3)e^{2\alpha-\beta}
-\chi^2\dif\phi_1 e^{2\alpha-\beta} +
\chi \dif (2\alpha-\beta) + \dif \chi \nonumber \\
A_{13} & = & -\chi e^{2\alpha-\beta} \label{fieldparam} \nonumber \\
A_{21} & = & (\phi_3 -\phi_1^2 + \dif\phi_1)
e^{2\alpha-\beta} \nonumber \\
A_{23} & = & e^{\alpha-2\beta} \\
\abar_{13} & = & \chi (\nu\phi_1-\mu) e^{\alpha-2\beta}
+ \nu e^{-\alpha-\beta} \nonumber \\
\abar_{23} & = & (\mu - \nu\phi_1) e^{\alpha-2\beta} \nonumber
\ea
Again, we want to construct an action $S$,
such that $A'$ and $\abar'$, defined
through $F(A,\abar)e^S\Psi=0 \Leftrightarrow
e^SF(A',\abar')\Psi=0$, are equal to
the connections $A^g$ and $\abar^g$
(the generalizations of (\ref{atra})
and (\ref{abartra})) that are the gauge
transforms with $g$ as in (\ref{defgg})
of the connections $A$ and $\abar$
mentioned in and below (\ref{defaa}). If $A^g=A'$
and $\abar^g=\abar'$ are satisfied,
$F(A',\abar')\Psi=0$ is equivalent to the statement
that $\Psi$ satisfies the $W_3$-Ward identities.
The connections $A'$ and $\abar'$ can be
obtained from (\ref{polara}) and (\ref{polarabar})
by first replacing $\var{X_i}$ by
$\var{X_i}+\vars{S}{X_i}$ everywhere,
followed by putting all terms in $\var{X_i}$ that
do not contain $\var{\mu}$ or $\var{\nu}$
equal to zero, as $\Psi$ depends only on $\mu$ and
$\nu$. Comparing these $A'$ and $\abar'$
with $A^g$ and $\abar^g$ yields a set of equations
for $\vars{S}{X_i}$, analogous to (\ref{sequ}).
These equations are necessary, but not sufficient,
because the $\var{\mu}$ and $\var{\nu}$
dependence of $A'$ and $\abar'$ must also be equal to
the $\var{\mu}$ and $\var{\nu}$
dependence of $A^g$ and $\abar^g$. Whether this is the case has
been verified by explicitly
computing the $\var{X_i}$ in (\ref{defpolara}) and (\ref{defpolarabar})
in terms of $\var{\mu}$, $\var{\nu}$,
and the functional derivatives with respect to the
fields in the Gauss decomposition (\ref{defgg}).
It turns out that this dependence is indeed
precisely the same.
The computations involved here are rather cumbersome, whether
there is a more direct way to see this,
is under current investigation \cite{jj}.
Due to this remarkable fact, we know
that if we now solve (\ref{Sequ1}) and (\ref{Sequ2}) for this case,
$Fe^S\Psi=0$ will be satisfied (with
this choice of parametrization and polarization) if and only if $\Psi$
satisfies the $W_3$-Ward identities.
Again, $S$ is given by (\ref{chiralaction}), and reads:
\ba
S & = & \deel{k}{2\pi}\int d^2z\,\,\Bigl[ -\hf
A^{ij}\dif \alpha_{i} \dbar
\alpha_{j}-\phi_i A^{ij}\dbar \alpha_{j}-\dbar\chi(\dif\phi_1
+\phi_1^2-\phi_3)e^{2\alpha_1-\alpha_2}\nonu
& &\hspace{1.3cm}+\mu((\dif-\phi_2)\phi_2 +\phi_2\phi_1-\phi_3)+
\nu(\dif-\phi_2)(\phi_3-\phi_2\phi_1)\Bigr],
\ea
where $\alpha_1=\alpha$, $\alpha_2=\beta$ and
$A^{ij}$ is the Cartan matrix of $Sl(3,\re)$,
$A^{ij}=\left( \begin{array}{rr} 2 & -1 \\ -1 & 2 \end{array} \right)$.
This action is important if one wants to
compute inner products of wave functions $\Psi$ in $Sl(3,\re)$
\cs . This will be the topic of the next section.

\newsection{The Inner Product and $W_3$ Gravity}

The wave functions $\Psi[\mu,\nu]=\exp S_W[\mu,\nu]$ that solve the
$W_3$-Ward identities, can be obtained from a constrained
Wess-Zumino-Witten model \cite{beroog,kj1}.
This means that at this stage we know the complete wavefunction
$e^S\Psi$. These wave functions solve the holomorphic $W_3$-Ward
identities, and can therefore be seen as
effective actions of chiral $W_3$-gravity. However, in
ordinary gravity there is a nontrivial coupling
between the holomorphic and the anti-holomorphic sectors, and
this is where part of the geometry of
two-dimensional quantum gravity comes in.
In this section we will consider the coupling between the
holomorphic and anti-holomorphic sectors of $W_3$-gravity, hoping
that it will lead to an understanding of
the geometry underlying the $W_3$-algebra. This
nontrivial coupling appears when computing inner products
of wave functions in $SL(3,\re)$ \cs .
The expression for such an inner product is
\be
\langle \Psi_1 \mid  \Psi_2 \rangle = \int D(\pri A )D(\prk\abar)
D(\prkd B) D (\prid \bar{B}) e^{V+S+\bar{S}}
\Psi_1[\mu,\nu] \bar{\Psi}_2[\bar{\mu},\bar{\nu}]. \label{innerprod}
\ee
The nontrivial coupling is due to the K\"{a}hler potential $V$, which
is associated to the symplectic form
defined by (\ref{eq:poisbracket}). To
find an expression for this K\"ahler potential, we first give the
definitions of $B$ and $\bar{B}$, the variables
on which the anti-holomorphic wave function $\bar{\Psi}_2$
depends. Let $H$ be the subgroup of $Sl(3,\re)$
consisting of all elements $h\in Sl(3,\re)$ satisfying
$h_{31}=h_{32}=0$;
$H$ can be conveniently parametrized by a Gauss decomposition.
Define the connection $B_0$ by requiring that $(B_0)_{31}=\bar{\nu}$ and
$(B_0)_{32}=\bar{\mu}$, and that $\prid F(B_0,\Lambda^t)=0$,
where $\Lambda^t$ is the transpose of
$\Lambda$. Then:
\ba
B & = & h B_0 h^{-1} - \dif h h^{-1}, \\
\bar{B} & = & h \Lambda^t h^{-1} - \dbar h h^{-1},
\ea
and the anti-holomorphic action $\bar{S}$ is given by
\be \bar{S} =
\deel{k}{2\pi}\int d^2z\,\,\tr(\prkd B \prk \bar{B})
+\deel{k}{2\pi}\int d^2z\,\,\tr(\Lambda^t h^{-1} \dif h )
-\Gamma_{WZW}[h].
\ee
In terms of $A$ and $B$, the K\"ahler potential is given by
\be
V=\deel{k}{2\pi}\int d^2 z \,\,
\tr (\pri A \prid \bar{B}- \prk{\abar} \prkd B).
\ee
The total exponent $K=V+S+\bar{S}$ occurring
in the inner product (\ref{innerprod}) is now a function of
$g$, $h$, and $\{\mu,\nu,\bar{\mu},\bar{\nu}\}$.
This "action" $K$ is part of the
covariant action of $W_3$-gravity.
The complete covariant action is given by
$K+S_W[\mu,\nu]+\bar{S}_W[\mub,\nub]$,
where $S_W[\mu,\nu]$ and $\bar{S}_W[\mub,\nub]$ are
the chiral actions for $W_3$-gravity
that were constructed in \cite{kj1}.
In the case of $Sl(2,\re)$, $K$ is equal to the Liouville action in a
certain background metric,
plus an extra term depending on $\mu,\bar{\mu}$ only. $K$ represents the
$W_3$-analogon of the Quillen-Belavin-Knizhnik anomaly.
Clearly, it will be interesting to have an explicit
expression for it. One can work out such an explicit expression,
by simply substituting all the expressions
given above. The result of all this is a large,
intransparant expression. However, upon further inspection,
it turns out that $K$ is invariant under
local $Sl(2,\re) \times \re$ symmetry transformations,
which can be used to gauge away four degrees of freedom.
Actually, one can proof (see \cite{jj}) that
$K$ only depends on the product $gh$. Using this
the action can be greatly simplified by
introducing a new Gauss decomposition for $gh$,
which is now an arbitrary element of $Sl(3,\re)$.
More specific, we take
\be
gh=\mat{1}{0}{0}{\rho_1}{1}{0}{\rho_3}{\rho_2}{1}
\mat{e^{\vp_1}}{0}{0}{0}{e^{\vp_2-\vp_1}}{0}{0}{0}{e^{-\vp_2}}
\mat{1}{-\rb_1}{-\rb_3}{0}{1}{-\rb_2}{0}{0}{1}.
\ee
Substituting this, one sees that
the action depends simply quadratically on $\ro_3,\rb_3$. Therefore,
ignoring subtleties arising from the measure
when changing variables from $A$ and $B$ to $gh$ and
$\mu,\nu,\mub,\nub$, we can perform the
$\int D\ro_3 D\rb_3 $ integration. The resulting action can be
written in the following form:
\ba
K = & &\hspace{-5mm}\deel{k}{2\pi}\int d^2 z\,\,\Bigl\{
       \hf A^{ij}\dif\vp_i\dbar\vp_j
      +\sum_i e^{-A^{ij}\vp_j}-A^{ij}(\rho_i+\dif \vp_i)
      (\bar{\rho}_j+\dbar \vp_j)\hspace{3cm}\\[.5mm]
  &   & \hspace{8mm}-e^{\vp_1-2\vp_2}(\mu-\hf \dif \nu -\nu\ro_1)
      (\mub+\hf \dbar \nub +\nub\rb_1)-e^{-\vp_1-\vp_2}\nu\nub\nonu
  &   & \hspace{8mm}
      -e^{\vp_2-2\vp_1}(\mu+\hf \dif\nu+\nu\ro_2)(\mub-\hf \dbar\nub
      -\nub\rb_2) +\mu T+\nu W + \mub \bar{T}
      + \nub \bar {W}\Bigr\}, \nonumber \label{finalres}
\ea
where we defined $T,W,\bar{T},\bar{W}$
through the following Fateev-Lyukanov \cite{fatly} construction:
\ba
(\dif-\ro_2)(\dif-\ro_1+\ro_2)(\dif+\ro_1) & = &
\dif^3+T\dif-W+\hf \dif T, \nonumber \\
(\dbar-\rb_2)(\dbar-\rb_1+\rb_2)(\dbar+\rb_1) & = & \dbar^3
+\bar{T}\dbar+\bar{W}+\hf \dbar \bar{T}, \label{fat}
\ea
and we shifted $\mu \rightarrow \mu -\hf \dif \nu$ and
$\mub\rightarrow\mub +\hf\dbar\nub$.
The first part of $K$ is precisely a chiral $Sl(3)$ Toda action,
confirming the suspected relation between $W_3$-gravity and Toda
theory. Actually, one would expect that in a "conformal gauge",
the covariant $W_3$-action will reduce to a
Toda action. Indeed, if we put $\nu=\nub=0$ in $K$,
then we find that $K$ is also purely quadratic in
$\ro_1,\ro_2,\rb_1,\rb_2$.
Performing the integrations over these variables as well, we find that
\be
K[\vp_1,\vp_2,\mu,\mub]=\deel{k}{4\pi}\int d^2 z \sqrt{-\hat{g}}
\left(\hf \hat{g}^{ab}\dif_a\vp_i\dif_b\vp_j A^{ij}+
4\sum_i e^{-A^{ij}\vp_j}+ R \vec{\xi}\cdot \vec{\vp}\right)
+K[\mu,\mub], \label{covtoda}
\ee
where $K[\mu,\mub]$ is the same expression as was
derived in \cite{herman}, namely
\be
K[\mu,\mub]=\deel{k}{\pi}\int d^2 z \,\, (1-\mu\mub)^{-1}
(\dif\mu\dbar\mub-\hf \mu (\dbar\mub)^2 -\hf \mub(\dif\mu)^2),
\ee
and $\hat{g}$ is the metric given by
$ds^2=|dz+\mu d\bar{z}|^2$. In the case of $Sl(3,\re)$,
$\vec{\xi}\cdot\vec{\vp}$,
with $\vec{\xi}$ being one half times the sum of the positive roots, is
just given by $\vp_1+\vp_2$.
The action (\ref{covtoda}) is the same Toda action that was originally
present in $K$ in a chiral form,
and the integration over $\ro_1,\rb_1,\ro_2,\rb_2$ has the effect of
coupling it to a background metric $\hat{g}$.

Of course, the most interesting part of the action is
the part containing $\nu,\nub$. Unfortunately, if we
do not put $\nu=\nub=0$, we can integrate over either
$\ro_1,\rb_1$ or over $\ro_2,\rb_2$, but not over
both at the same time, due to the presence of third order terms in $K$.
Another clue regarding the contents
of the action (\ref{finalres}) can be obtained by treating the
second and third line in (\ref{finalres}) as perturbations
of the first line of (\ref{finalres}).
This means that we try to make an expansion in terms of
$\mu,\mub,\nu,\nub$.
The saddlepoint of the $\rho$-terms is at $\ro_i=-\dif\vp_i$
and $\rb_i=-\dbar\vp_i$.
{}From (\ref{fat}) we can now see that $T,W,
\bar{T}, \bar{W}$ are, when evaluated in this saddle point,
the (anti)holomorphic
energy momentum tensor and $W_3$-field that are
present in a chiral Toda theory
\ba
T &=& -\hf A^{ij}\dif \vp_i \dif \vp_j - \vec{\xi} \cdot \dif^2
\vec{\vp}, \nonu
W &=& -\dif \vp_1 ((\dif \vp_2)^2+\hf \dif^2\vp_2-\dif^2\vp_1)
+\hf\dif^3 \vp_1 - (1\leftrightarrow 2),
\ea
and similar expressions for $\bar{T},\bar{W}$.

This suggests that the full
action $K$ contains the generating functional for
the correlators of the energy-momentum tensor and the
$W_3$-field of a Toda theory, "covariantly"
coupled to $W_3$-gravity. The presence of the third order terms in
$W,\bar{W}$ in (\ref{finalres}) prevents us from computing the
action of this covariantly coupled Toda theory.

Detailed proofs, that were omitted here,
as well as generalizations to other $W$-algebras,
will be the subjects of a future publication \cite{jj}.

This work was financially supported by the
Stichting voor Fundamenteel Onderzoek der Materie
(FOM).


\newpage

\begin{thebibliography}{99}
\bibitem{herman} H. Verlinde, \np{337} (1990) 652.
\bibitem{sammy} A. B. Zamolodchikov, Theor. Math. Phys, {\bf 65}
(1985) 1205.
\bibitem{beroog} M. Bershadsky and H. Ooguri, \cmp{126} (1989) 49.
\bibitem{ber} M. Bershadsky, \cmp{139} (1991) 71.
\bibitem{moore} S. Elitzur, G. Moore, A. Schwimmer and N.
Seiberg, \np{326} (1989) 108.
\bibitem{poly} A. M. Polyakov, \intmod{5} (1990) 833.
\bibitem{witten1} E. Witten, \cmp{137} (1991) 29.
\bibitem{witten2} E. Witten, `On Holomorphic Factorization of
WZW and Coset Models,' IASSNS-preprint, IASSNS-HEP-91/25 (June
1991).
\bibitem{jj} J. de Boer and J. Goeree, in preparation.
\bibitem{bilal1} A. Bilal, V. V. Fock and I. I. Kogan, `On the
Origin of W-Algebras,' CERN-preprint, CERN-TH 5965/90 (December
1990).
\bibitem{bilal2} A. Bilal, `W-Algebras from Chern-Simons
Theory,' CERN-preprint, CERN-TH 6145/91, LPTENS 91/17.
\bibitem{kj1} H. Ooguri, K. Schoutens, A. Sevrin and P. van
Nieuwenhuizen, `The Induced Action of $W_3$ Gravity,' preprint,
ITP-SB-91/16, RIMS-764 (June 1991).
\bibitem{kj2} K. Schoutens, A. Sevrin and P. van
Nieuwenhuizen, \np{349} (1991) 791, \plb{243} (1991) 248.
\bibitem{pope} E. Bergshoeff, C. N. Pope, L. J. Romans,
E. Sezgin, X. Shen and K. S. Stelle, \plb{243} (1991) 330.
\bibitem{hull} C. M. Hull, \np{353} (1991) 707.
\bibitem{fatly} V. Fateev and S. Lukyanov, \intmod{3} (1988)
507.
\end{thebibliography}
\end{document}